# Generative design of inorganic materials

Jose Recatala-Gomez[1,2], Haiwen Dai[3], Ruiming Zhu[3], Nikita Kazeev[4], Wei Nong[3], Gang Wu[5], Maciej Koperski[4,6], Tan Teck Leong[5], Andrey Ustyuzhanin[4,7,8], Gerbrand Ceder[1,9], Kostya Novoselov[4,6], Kedar Hippalgaonkar[2,3,4#]

[1] Department of Materials Science & Engineering, University of California, Berkeley, California 94720, United States

[2] Berkeley Education Alliance for Research in Singapore (BEARS), 1 Create Way, #05-02/03, CREATE Tower, 138602, Singapore

[3] School of Materials Science and Engineering, Nanyang Technological University, 50 Nanyang Avenue, Singapore 639798

[4] Institute for Functional Intelligent Materials, National University of Singapore, Singapore 117544, Singapore

[5] Institute of High-Performance Computing, Agency for Science, Technology and Research, 1 Fusionopolis Way, #16-16 Connexis, Singapore 138632, Singapore

[6] Department of Materials Science and Engineering, National University of Singapore, Singapore 117575, Singapore

[7] Constructor Knowledge Labs, Campus Ring 1, Bremen 28759, Germany

[8] Constructor University, Campus Ring 1, Bremen 28759, Germany

[9] Materials Sciences Division, Lawrence Berkeley National Laboratory, Berkeley, California 94720, United States

[#]Email: kedar@ntu.edu.sg

## Abstract

Materials discovery is fundamental to advance next-generation technologies as well as for sustainable and circular economy. Beyond computational screening, generative models are efficient at finding materials with desired properties, *via* multi-modal learning using multiscale data. This perspective examines the landscape of generative design for inorganic materials and discusses the integration of multi-modal learning with high-throughput experimental validation. We contextualize these challenges through the lens of a generative design framework as a unified approach to address the data-driven inverse design of functional materials. The central idea of the framework is constructed around a foundation AI model for inorganic materials interlinked deeply with various property databases and high-throughput experiments *via* a machine learning driven closed loop, which enables the framework to solve key challenges in functional materials. We argue that domain-specific implementations of such integrated workflows represent a promising pathway toward the unresolved challenge of data-driven inverse design for atom-engineered inorganic functional materials.

## Introduction

Materials science is undergoing a revolution due to artificial intelligence (AI), the acceleration of computation, and automation. The 2024 Nobel Prizes in Physics and Chemistry were awarded to works leveraging AI to solve key scientific problems. In inorganic materials, however, significant challenges remain due to the complexity of crystal structures, defects, and the need to accurately represent physical properties across multiple length scales. Recent advances in machine learning have enabled predictive breakthroughs[1]: for instance, a Cu-Al catalyst achieved ~80% Faradaic efficiency for $CO_2$-to-ethylene conversion versus ~66% for pure Cu[2]; an autonomous laboratory synthesized 41 compounds in just 17 days.[3] Machine Learning approaches now predict key properties like energy and band gap with near-DFT accuracy when trained on ~$10^4$-$10^5$ inorganic crystalline structures.[4] Still, truly generative design is yet to be validated. Such a design methodology should go beyond the search of pure materials, but extend to doped crystals, alloys, as well as pay attention to multi-scale property-directed materials design, where materials properties can be modified by their composition, crystal structure, dopants, morphology, etc.

In 2023, Merchant *et al.* introduced GNoME, which leveraged graph neural networks to expand the known materials landscape by predicting approximately 380,000 thermodynamically stable crystal structures.[5] Building on this foundation, the first instance of GenAI-enabled discovery of a new material was achieved by Zeni *et al.*[6] They developed MatterGen, a diffusion-based model that generated stable materials, a disordered version of one of the predicted compounds ($TaCr_2O_6$) which was experimentally realised and matched the predicted bulk modulus within 20%. Large-scale Machine Learning screening of ~32 million candidates coupled to high-throughput experiments performed via expert selection delivered new mixed cation $Na_xLi_{3-x}YCl_6$ ($0 \leq x \leq 3$) solid electrolytes with improved ionic transport.[7] Despite these milestones, similar studies are still nascent due to significant conceptual bottlenecks, which remain unresolved. An important bottleneck arises from the strict constraints governing crystal formation, which originate from factors such as the symmetry of the crystal structure, atomic orbitals, and the nature of chemical bonding. In addition, compositional disorder and structural imperfections can reduce the energy of formation, yet these effects remain difficult for models to learn due to limited data and are therefore often better encoded through descriptor engineering.[8] Beyond this, current generative models are limited to perfect, undoped crystals, restricting its use in discovering, new, high performing functional materials. Hence, extending such AI-driven methods to explore doped crystals, disordered systems and alloys is essential for designing functional materials with tunable properties. Addressing these challenges requires the integration of three key emerging capabilities:

(i) physics-aware representations that encode crystallography, symmetry, and defects,

(ii) generative models capable of exploring compositional and structural space under property constraints, and

(iii) validation pipelines that connect computational predictions with experimental realization.

It appears that the need of the hour is an integrated approach driven by a generative design framework shown in Figure 1. This framework integrates symmetry- and defect-aware AI models with autonomous experimental workflows. Firstly, materials with target properties are designed *via* a generative model, with structure-property relationships learnt from existing computational and experimental knowledge. Principles of crystallography are injected into the model training, including symmetry operations, partial disorder, and space group permutations, to provide a more complete structural representation and thus offer a higher level of understanding between crystals of different space groups, local coordination environment and motifs. The inclusion of fundamental crystallographic and orbital-pair bonding knowledge largely reduces the likelihood of generating "unsensible" materials, potentially caused by the lack of negative samples (*i.e.* unrelaxed theoretical structures, not-yet made materials) and training data bias. Fine tuning of the band structure and stability can further be achieved by a seamlessly integrated defect-aware model pre-trained on domain-specific datasets as well as embedding learnt from Machine-Learned Interatomic Potentials (MLIPs).[9]

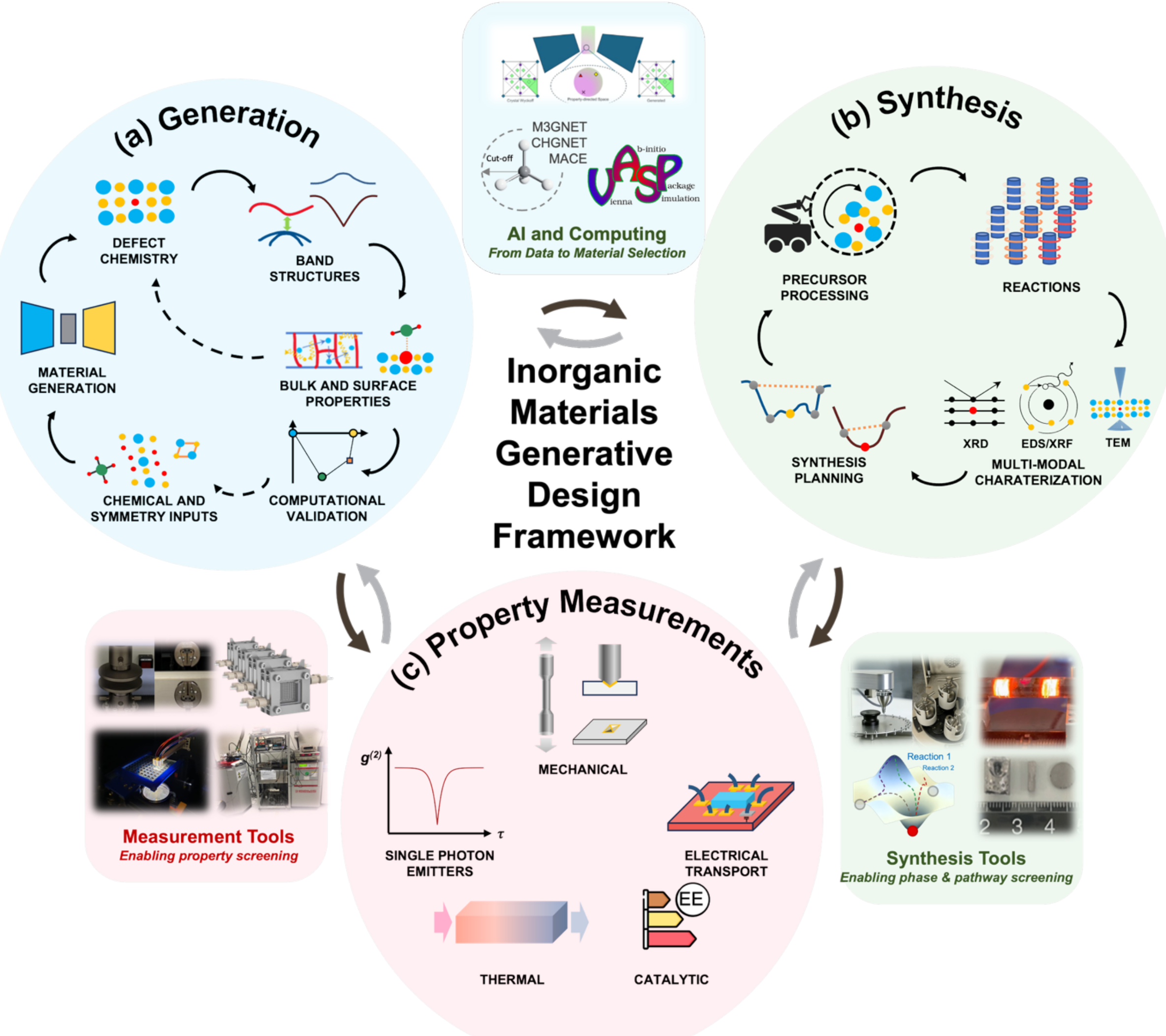

**Figure 1. Schematic representation of the end-to-end generative design framework**. Generative design includes property-directed machine learning models and *in-silico* validation using DFT. Only when 'confident' with the predictions (structure, bonding, stability, decomposition etc.), the predicted compound undergoes experimental validation using data-driven high-throughput synthesis and characterization, including not just materials properties (e.g. structure, composition) but also functional property testing (e.g. mechanical).The seamless integration of state-of-the-art Machine Learnt Interatomic Potentials (MLIPs), with DFT and MD, along with rapid experimental testing of structure, composition as well as desired properties will help to bridge the experiment-theory gap.

Following generation, candidate materials need to be evaluated using high-throughput simulations and subsequently passed to materials acceleration platforms (MAPs) or self-driving laboratories (SDLs) for experimental validation. Synthesis planning and process optimization locks the correct recipe with the highest yield and selectivity. Successful synthesis with a correct structure and high phase purity is further characterized to validate its physical properties. Validation results, especially structure-property deviations, are provided as feedback to actively refine the generative model, to stepwise close the potential gaps between generation and validation, forming a true closed-loop property-driven autonomous material discovery framework. In the near future, such workflows could be partially or fully run by AI agents, including physical AI (integrating into the physical MAPs/SDLs). In this perspective, we discuss the conceptual components required for an integrated generative design framework, situate them within the broader landscape of emerging approaches, and conclude with a perspective on future challenges and a Materials-in-Context box summarising technological maturity, key challenges and potential impact.

## Components of the generative design framework

Generative design of inorganic materials can be understood as the integration of three closely linked components: i) physics-aware materials representation, ii) domain-agnostic generative design towards the development of a foundation model and iii) a feedback loop via in-silico and experimental validation. While different frameworks may implement these components in distinct ways, their combination appears essential for realizing practical inverse design workflows. In the following, we discuss each component and illustrate how they can be integrated within approaches such as the generative design framework shown in Figure 1.

### *Inverse Design*

Property-directed inverse design can broadly be approached through two paradigms: **high-throughput screening** and **generative design**. *High-throughput screening* can be achieved by scanning the full chemical space and calculating every possible combination of composition and structure with every property that is desired for a particular application. Although computational methods and MLIPs have enabled the evaluation of millions of candidate structures, exhaustive screening remains intractable for two key reasons: (1) calculating each desired property for a myriad of combinations spanning the chemical space is intractable, (2) many properties required a hierarchical featurization, which should cover not just a single phase of a material (typically computed by first-principles theory), but instead doped, defective and multi-phase compounds and their polymorphs. Hence, multiscale calculations spanning across atomic, microstructure and even structural domains are necessary to cover the vast inverse design space.

In contrast, *generative design* aims to directly propose candidate materials conditioned on target properties, offering a more data-efficient and potentially scalable alternative. In addition, such generative models can be embedded in closed loop experimentation through information-gain driven optimization workflows, making material discovery feasible. However, current generative approaches remain limited in their ability to reliably produce physically realistic, synthesizable materials, highlighting the need for tighter integration with physics-based models and experimental validation.

A practical generative design framework requires the integration of multiple modeling layers, including: (1) models for material representations, (2) machine-learning enabled interatomic potentials (MLIPs) or force fields (MLFFs) which are trained on vast variety of simulated and experimentally characterized materials (Materials Project[10], Alexandria[11], JARVIS[12], OQMD[13], ICSD[14]) and are specialized to predict atomic energy and forces given material's atomic structure, (3) case-specific conditional generative models generative models. Conditional generative models should map from a latent space to a reconstructable material representation that is used to obtaining atomic structures with desired property.[15] These components are increasingly being developed across the community and are beginning to converge toward unified, multi-scale generative workflows.

Currently, MLIP models, such as MACE[9], MEGNet[16], CHGNet[17], and UMA[18] have demonstrated broad applicability across materials systems, particularly for evaluating static properties such as stability and energetics of given atomic configurations.[19,20] However, they primarily operate in a forward-prediction regime and lack the inherent capability to generate novel structures under property constraints. [13,14]Moreover, MLIPs are typically trained on near-equilibrium structures and may struggle to capture complex energy landscapes, metastable phases, or highly disordered systems. This fundamental limitation requires either integration of MLIPs with dedicated generative architectures that can sample the chemical and structural space intelligently or extension of these to predict semantic material intermediate representation that can be used across a variety of generative tasks like quantum materials, high-entropy materials for electrocatalysis, or thermal materials.

### Inorganic Materials Representation/Tokenization

The success of generative design models critically depends on how materials are represented in a machine-readable form. These representations are often referred to as descriptors, fingerprints, or embeddings and they translate complex structural and compositional information into formats suitable for learning algorithms (Figure 2).[21] These numerical translations are variously referred to as representations, tokens, descriptors, fingerprints, or features. The choice of representation (Table 1) is just as important as the choice of the ML algorithm itself for ability to learn meaningful patterns, generalize to unseen data, and ultimately, enhance predictive or generative power. Designing effective representations remains an open challenge, requiring trade-offs between model efficiency, physical interpretability, and the ability to reconstruct atomic structures. No single representation currently satisfies all desirable properties simultaneously, and the choice of representation can strongly influence both predictive and generative performance.

**Table 1. Different representation types with key characteristics**. Different representations are ranked by their reconstructability, from impossible to easy. The colour indicates whether the suitability of the representation in forward tasks, with red being limited, green being high and blue being moderate.

| Representation Type | Key Characteristics | ML-Friendliness | Reconstructability | Inductive Bias | Suitability (Generative) | Examples |
|---|---|---|---|---|---|---|
| **Compositional** | Elements, Stoichiometry; Low Dim.; Invariant | (+) Simple, Cheap; (-) Lacks structure | Impossible | Chemical rules | No (cannot generate structure) | MagPie, ElemNet |
| **Learned Embeddings** | Varies (Composition, Structure); Low-Med Dimensions; Varies (can be learned) | (+) Potentially optimal for task, Captures subtle patterns; (-) Black box, Data hungry | Difficult/Impossible (unless designed, e.g., VAE) | Learned from data (implicit) | Moderate (requires invertible decoder, e.g., VAE) | VAE latent space, GNN node embeddings |
| **Local Structural Descriptors** | Atomic Coordinates (local env.); Med-High Dim.; Invariant (ACSF, SOAP) | (+) Captures local geometry, Transferable; (-) Param tuning, May miss long-range | Difficult/Ambiguous (assembling local envs.) | Locality, Symmetry (if built-in) | No (cannot generate structure) | ACSF, SOAP |
| **Global Structural Descriptors** | Atomic Coords (full structure); High Dim.; Varies (e.g., CM invariant, RDF not) | (+) Captures global structure; (-) Costly for large systems, May lack local detail | Varies (e.g., RDF hard, CM needs sorting) | Many-body interactions, Symmetry (if built-in) | Varies (depends on invertibility) | Coulomb Matrix, RDF, ACE, MTP |
| **Structural: Graph-based** | Atoms (nodes), Bonds/Neighbors (edges); Variable Dim.; Equivariant (w/ E-GNNs) | (+) Natural connectivity, Powerful w/ GNNs; (-) Edge definition, | Varies (inferring 3D from graph, depending on encoding methods) | Connectivity, Locality, Symmetry (E-GNNs) | No (cannot generate structure) | CGCNN, M3GNet, NequIP, MACE |

| | | Graph isomorphism | | | | |
|---|---|---|---|---|---|---|
| **Structural: Voxel/Grid-based** | Atomic Coords mapped to grid; Very High Dim.; Not inherently invariant | (+) Represents 3D fields, Suits CNNs; (-) High dim., Resolution-dependent, Invariance handling | Easy (direct mapping) | Spatial occupancy | Yes | iMatGen |
| **Structural: Symmetry-based** | Asymmetric unit + Ops / Wyckoff positions; Low-Med Dim.; Explicit Symmetry | (+) Compact, Enforces symmetry; (-) Complex implementation | Easy (defined procedure) | Crystallographic symmetry | Yes (ensures symmetry by construction) | WyCryst, WyFormer SymmCD |

Effective material representations must balance three key aspects: ML-friendliness, reconstructability, and inductive bias. However, these objectives often diverge, and existing approaches typically optimize for only a subset of these criteria. ML-friendliness ensures that representations maintain a smooth, learnable relationship between representation and target properties. Reconstructability refers to the ability to uniquely recover the full crystal structure from a representation. High-level structural representations simplify model learning but hinder reconstruction, while atomic coordinates enable exact reconstruction but demand learning complex atom interactions and symmetries. Inductive bias embeds prior material knowledge, improving representation quality. Among these prior knowledge, MLIPs pr periodicity ensures consistent representation for periodic crystals, while symmetry captures invariances to translations, rotations, and permutations, crucial for scalar and tensorial property predictions. Finally, chemical similarity refers to the degree to which materials resemble each other based on their encoded structural, compositional, or property-related features. No representation perfectly balances these aspects: handcrafted descriptors[22] offer ML-friendliness and inductive bias but lack reconstructability, while direct coordinate representations are fully reconstructible but require models to handle symmetries, illustrating the challenge in designing optimal representations.

While representations of ideal, bulk crystalline materials are mature[15], modelling features such as defects, disorder, surfaces, and interfaces poses added complexity due to broken periodicity and symmetry. Point defects (e.g., vacancies, interstitials, substitutions) are often captured by modifying host lattice representations, such as altering graph node features[15] or using relative difference features.[23] Advanced methods like DefectNet use universal graphs with defect markers to learn defect-specific interactions.[24] Extended defects (e.g., dislocations, grain boundaries) require large simulation cells and are often represented via local descriptors like Smooth Overlap of Atomic Positions (SOAP)[25], the scattering transform.[26] Grain boundaries can be described by macroscopic geometric parameters such as misorientation or through graph-based methods connecting atoms across the interface.[15] Amorphous materials, which lack long-range order, are best described using local[27] or statistical descriptors, including topological data analysis[28] or pair density matrices[29], to capture short-range order. Surfaces and interfaces are typically modelled with slab geometries in 2D[30] and represented using features like Miller indices, surface coordination, and adsorbate properties.[31] Graph-based methods and electronic structure descriptors (like density of states or d-band center) are also employed.[15] Generative models for surface structures are emerging, needing to account for substrate registry and surface reconstruction.[32] Dis-GEN[8] introduces a symmetry-aware generative model that enables the generation of inorganic crystal structures with compositional disorder and partial occupancies, trained on the International Crystal Structure Database (ICSD).[33] Representing these complex, non-ideal structures effectively, especially for generative tasks aiming to design materials with specific defects or interfaces, remains a significant frontier in property-directed materials design, particularly for applications where such features dominate functional behaviour.

## Machine Learning Interatomic Potentials (MLIPs) as a precursor to Materials Foundation Models

Atomistic simulations in materials science depend on potential energy surface models, balancing speed and transferability in empirical force fields with the accuracy and computational cost of quantum methods like DFT. MLIPs have become a central tool in computational materials science, bridging the gap between the accuracy of quantum mechanical methods and the efficiency of classical force fields. They have emerged over the past 15 years and bridge the gap between the speed of classical potentials and the accuracy of DFT. MLIPs use flexible machine learning models, typically neural networks or kernel methods, trained on large datasets of DFT-calculated energies and forces for various atomic configurations. Once trained, these models can predict energies and forces with near-DFT accuracy but at a computational cost orders of magnitude lower, approaching that of classical potentials. This enables large-scale, long-time atomistic simulations with unprecedented fidelity.[34] Despite their rapid progress, MLIPs face ongoing challenges related to transferability, long-range interactions, and robustness under extrapolation, particularly in chemically complex or defect-rich systems. In addition, they can only be as accurate as DFT, so they inherent the same problems in linking to experimental validation that computational approaches have suffered from in the last few decades.

The first generation MLIPs used handcrafted, physics-informed descriptors of local atomic environments, ensuring invariance to rotation, translation, and permutation (Behler-Parrinello[35] and Gaussian Approximation Potentials).[36] The second generation introduced end-to-end learning via Graph Neural Networks (GNNs), replacing manual descriptors with learned representations. MEGNet[37] pioneered invariant message passing, while M3GNet extended this by incorporating three-body interactions, enabling broader chemical coverage and higher fidelity.[38] The third generation introduced equivariant GNNs, which explicitly incorporate rotational equivariance. Some architectures exemplifying this approach include PaiNN[39], NequIP[40], Allegro[41], and MACE[42], offering substantial gains in accuracy and property prediction. Most recently, fourth-generation models have adapted Transformer-based architectures (e.g., Equiformer[43], Graphormer[44]), using attention mechanisms to better capture long-range interactions and scale to large systems. To address limitations in modelling electrostatics and dispersion, hybrid methods like latent Ewald summation have been proposed to embed long-range physics into MLIPs, improving reliability for charged or polar systems.[45] While such MLIPs significantly accelerate the evaluation of energies and forces of generated crystals, they do not completely replace DFT in benchmarking and validating due to prediction inaccuracy and softening effects[46], as well as degraded extrapolation. Addressing these limitations requires a community effort to develop better evaluation metrics, improve model design, and, crucially, enhance the quality and quantity of training data.

### *Generative Design of Inorganic Materials*

Generative models offer a promising route for exploring previously inaccessible regions of materials space by proposing new structures and compositions conditioned on desired functionality. Such generative models have produced varying success in computer-assisted drug discovery, particularly for the *de novo* design of molecules with desired properties. At the core of the generative design framework is the development of a generative model for inorganic materials, trained on extensive datasets of material properties to predict new materials with specific functionalities. Building such a model for inorganics entails different challenges relative to those for organics due to limited high-quality data, as well as the necessity to maintain the crystallographic symmetries and various complex bonding energies and co-ordination environments. Inorganic materials require specialized techniques like X-ray diffraction (XRD) and electron microscopy to capture their structure-property relationships, thus needing advanced feature engineering and domain-specific expertise to incorporate these complex datasets into AI models. A robust foundation model can accelerate the discovery of materials even effectively replacing or augmenting the standard, yet expensive DFT calculations for crystal structure prediction (CSP) tasks.

| Generative Design Algorithm Task | Description | Examples |
| --- | --- | --- |
| **De Novo Generation** | Generate entirely new crystal structures that are physically plausible, typically within a learned distribution of symmetry, stability, and compositional validity. | WyckoffDiff [47], WyFormer [48] |
| **Crystal Structure Prediction (CSP)** | Given a chemical composition, predict the lowest-energy crystal structure or all plausible structures that can exist at equilibrium. | DiffCSP, [49] DiffCSP++[50] |
| **Metastable State Generation** | Generate non-ground-state structures for a given composition that satisfy symmetry constraints (e.g., specific space group), useful for applications like piezoelectrics. | PGCGM[51] |
| **Inverse Design / Conditional Generation** | Given a desired property or function (e.g. band gap, formation energy, thermal conductivity), generate candidate materials conditioned on that target, often via guided sampling or reinforcement learning. | WyCryst[52], FTCP[53], Chemeleon2[54], CrystalFormer[55], |
| **Template-Based Exploration** | Enumerate new structures based on symmetry or Wyckoff templates, possibly varying composition or local environment within fixed structural motifs. | SymmCD[56], Plaid++[57] |

Initial efforts in generative models for crystal structures used Variational Autoencoders (VAEs)[58] and Generative Adversarial Networks (GANs)[59], which learned latent representations of materials and decoded them into candidate structures. Diffusion models have since become dominant, offering stable training, high-quality outputs, and a natural fit for incorporating physical priors.[60] Key innovations include CDVAE[61], which combines a VAE encoder with a diffusion decoder operating *via* equivariant GNNs, and DiffCSP[62]/DiffCSP++, which diffuse fractional coordinates and lattice parameters while incorporating space group constraints.[63] MatterGen further extends diffusion methods by operating directly on atom types, Cartesian coordinates, and lattice vectors, enabling conditional generation across diverse material properties.[6]

Other paradigms have emerged as well. Flow-based models like CrystalFlow[64] and FlowMM[65] use continuous normalizing flows to transform simple base distributions into structured crystal configurations, often relying on equivariant GNNs and Riemannian flow matching for improved fidelity. Autoregressive models and large language models (LLMs) offer complementary strengths by capturing high-level chemical patterns through sequential generation. Hybrid models like WyFormer[64,65] and FlowLLM[65] combine autoregressive generation of abstract structural tokens (e.g., Wyckoff positions) with geometric refinement via diffusion, MLIPs, and flow models. Another autoregressive framework, CrystalFormer, models the joint probability distribution over species, Wyckoff positions, and atomic coordinates, enabling both property-directed inverse design and crystal structure prediction through tailored architectural variants.[55,66,67] This modular strategy improves scalability and diversity while respecting crystallographic symmetry, marking a promising direction for generative materials discovery. Chemeleon2[54] integrates a VAE-based latent representation of crystal structures with a denoising diffusion model acting as the policy within a reinforcement learning (RL) framework, where reward-guided optimization enables property-driven inverse design and possibly de novo generation.

The combination of generative models for structure/composition suggestion and MLIPs for rapid property evaluation creates a powerful synergy: generative models explore the vast materials space guided by property objectives, while MLIPs provide computationally efficient validation of thermodynamic feasibility and basic property estimates, filtering candidates before expensive DFT calculations or experimental synthesis.

Integrated generative design frameworks are often structured as multi-stage pipelines that connect representation learning, property prediction, and inverse design. One example is the three-stage pipeline illustrated in Figure 2, which we describe here to highlight a possible implementation.

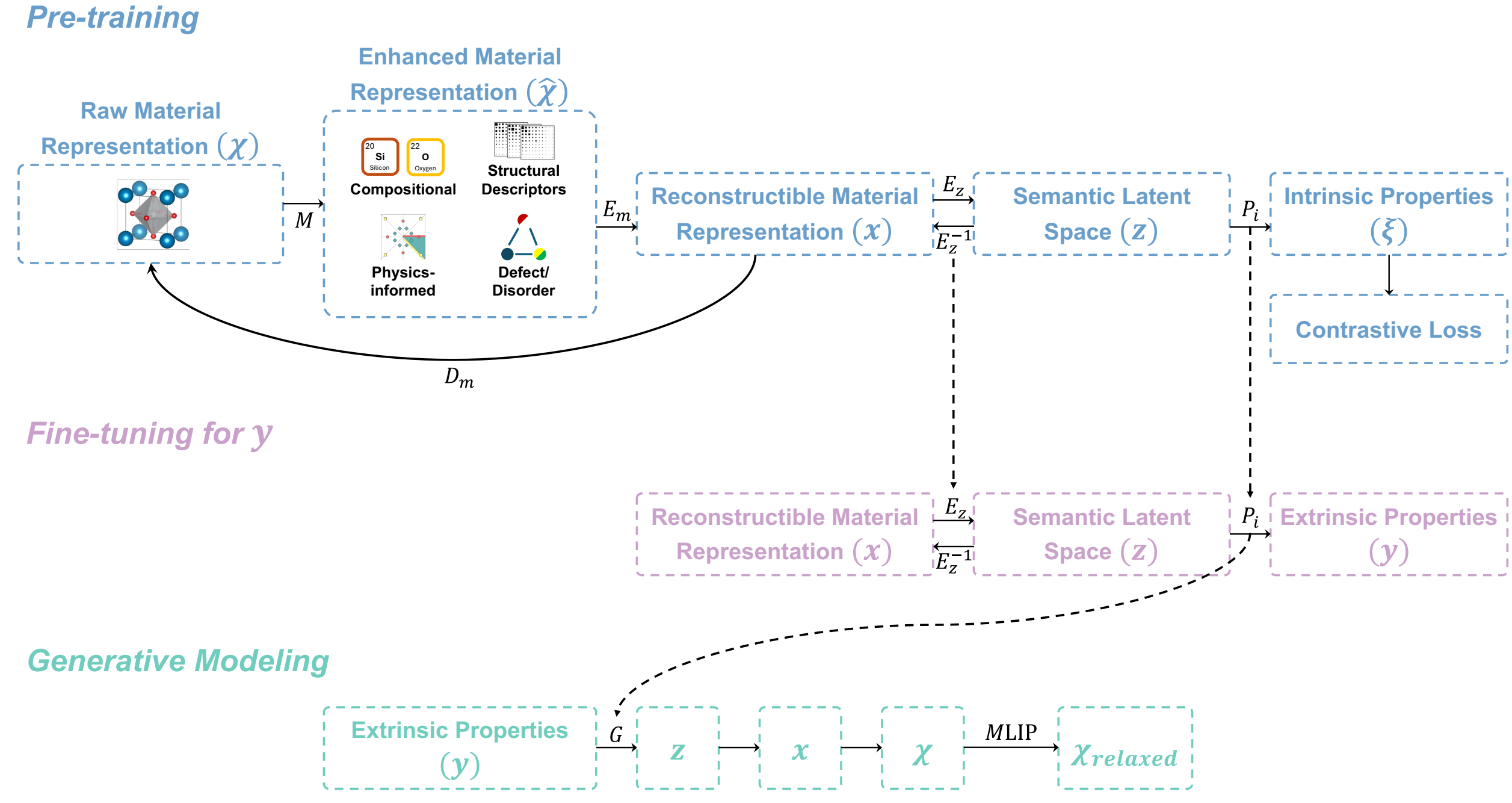

**Figure 2. Proposed Generative Design Framework Pipeline Architecture.** Pretraining (upper), Fine-tuning (middle), Generative training and inference (bottom).

### *Stage 1: Foundation Model Pretraining*

The foundation model pretraining stage establishes robust material representations capable of capturing the complex relationships between structure, composition, and properties. Starting with raw material representations ($\chi$), the system creates enhanced material representations that integrate multiple complementary descriptors: defect/disorder representations for capturing structural imperfections, physics-informed bulk representations encoding crystallographic principles, compositional descriptors for chemical information, and structural descriptors for geometric features.

These enhanced representations are then processed through an encoder $E_m$ to create reconstructible material representations (x), which are subsequently mapped to a semantic latent space (z) *via* encoder $E_z$. The semantic latent space serves as a unified representation that captures the essential features governing material behavior. From this latent space, intrinsic properties (ζ) are predicted through predictor $P_i$, and the system is trained using contrastive loss to ensure that materials with similar properties have similar latent representations while maintaining the ability to reconstruct the original material structure through decoder $E_z^{-1}$.

A critical prerequisite for the proposed pretraining stage is a sufficiently large and diverse corpus of structure-property data. Existing open databases: Materials Project (~154k inorganic compounds), OQMD (~1M entries), Alexandria (~4.5M), and ICSD (~280k experimental structures), collectively provide on the order of $10^5$-$10^6$ bulk crystalline entries, predominantly for ideal, defect-free stoichiometric phases. By contrast, systematically computed defect data (formation energies, charge transition levels, migration barriers) exists only for narrow slices of chemical space, typically a few hundred host-defect combinations per study. This creates a severe data bottleneck for the defect-aware foundation model central to the generative design framework: contrastive pretraining requires paired structure-property examples at a scale that defect databases cannot yet supply. We recommend that practitioners quantify the minimum pretraining corpus size they anticipate, by analogy with vision and language foundation models, where effective representation learning typically requires $10^6$-$10^8$ examples, and discuss concrete strategies for closing this gap. Promising directions include multi-fidelity training pipelines that combine cheap MLIP-relaxed defect structures with sparse high-fidelity DFT references, active-learning loops that prioritize informative defect configurations for expensive calculations, and synthetic data augmentation via symmetry operations and compositional interpolation across known host lattices. Without such a discussion, the feasibility of Stage 1 remains uncertain, and the three-stage pipeline risks appearing aspirational rather than actionable.

### *Stage 2: Task-Specific Fine-Tuning*

The fine-tuning stage adapts the pretrained foundation model for specific applications and target properties. Using the reconstructible material representation (x) as input, the system leverages the frozen, pretrained encoder $E_z$ to map materials to the semantic latent space (z). The key innovation in this stage is the training of a new property predictor $P_f$ that maps from the latent space to extrinsic properties (y), the specific functional properties required for target applications such as catalytic activity, thermal conductivity, or electronic band gaps.

This transfer learning approach ensures that the rich material representations learned during pretraining are preserved while enabling rapid adaptation to new property prediction tasks. The frozen encoder maintains the fundamental understanding of material structure-property relationships, while the new predictor learns domain-specific mappings relevant to the application at hand.

### *Stage 3: Generative Design and Structural Optimization*

The generative design stage implements inverse materials design by conditioning material generation on desired properties. Given a target extrinsic property (y), a generative model G is trained to map from property space to the semantic latent space generates appropriate latent representations (z). These latent representations are then decoded using the pretrained decoder $E_z^{-1}$ to produce reconstructible material representations (x).

The generated representations are subsequently converted to full atomic structures (X) with explicit atomic coordinates, compositions, and lattice parameters. To ensure physical realism and structural stability, the generated structures undergo relaxation using MLIPs and if-needed DFT, producing optimized structures ($X_{relaxed}$) that maintain the target properties while satisfying thermodynamic and crystallographic constraints.

This three-stage pipeline creates a complete inverse design workflow that progresses from property specifications to validated material structures, enabling systematic exploration of materials space guided by functional requirements while maintaining physical and chemical feasibility.

**Core Challenges in Crystal Generation**

Generative modeling for inorganic crystals has made significant strides, but several challenges remain. A critical issue exists in structure generation attempts: sometimes, sampled crystal structures may not be physically meaningful upon reconstruction. This stems from the fact that crystallography does not guarantee physical meaning, only ensures symmetry-consistent possibilities and respecting local and global bonding rules. This is the primary reason why “random” (unstable) structures are produced without 100% validity and most generated structures (which tend to be stable) undergo non-trivial DFT relaxations (undergoing symmetry and/or bonding change). These challenges highlight a fundamental gap between generative capability and physical validity, which remains an outstanding challenge in the field.

Stability is another key issue, as generated materials must be stable under normal conditions. Current stability estimation methods, such as energy above the convex hull, produce varying results (5-35% thermodynamically stable novel structures)[64,65], and their correlation with real-world stability is uncertain due to limitations in DFT calculations.[68,69] Accurately capturing crystallographic symmetry is also crucial for realistic property prediction. While symmetry-compliant models succeed by ensuring control over symmetry in the generation process, other models often struggle to produce realistic structures that would be considered credible by experimentalists. Most importantly, generated crystal structures need to be synthesized (different from stability) and until now, are precluded from experimental validation and real-world application. Thermodynamic stability alone is insufficient, as factors like kinetics and precursor availability are often not captured in current models.[70–73] Existing methods for predicting synthesizability are unreliable, and integrating accurate synthesizability prediction into the generative process remains an unsolved problem, creating a gap between computational predictions and experimental realization.

Bridging this synthesizability gap requires moving beyond thermodynamic (meta)stability as the sole filter for generated candidates. We recommend a concrete synthesizability-aware strategy, addressing at least three complementary directions. First, retrosynthetic feasibility scoring approaches such as Retro-Rank-In[74] learn to rank hypothetical compounds by the likelihood that a viable synthesis route exists, trained on positive examples from ICSD and negative examples from computationally stable but never-synthesized entries in Materials Project. Integrating such a scorer as a post-generation filter in Stage 3 could significantly reduce the fraction of "paper crystals" passed to experimental validation. Second, encoding human synthesis knowledge extract heuristic rules[75] (common precursors, accessible oxidation states, known structural prototypes) from the experimental literature and embed them as soft constraints or auxiliary loss terms during generative training; this biases the model toward compositions and structures that practicing chemists would recognize as plausible. Third, kinetics-aware screening (thermodynamic stability methods such as energy above the convex hull) does not capture whether a material can actually nucleate and grow under accessible conditions; incorporating kinetic descriptors such as nucleation barriers, metastable phase lifetimes from MLIP-driven molecular dynamics, or synthesis-temperature-dependent phase diagrams would add a crucial reality check that current generative frameworks uniformly lack. Together, these directions would transform the current "generate-then-hope" pipeline into one where synthesizability is a necessary design objective alongside target properties a natural fit for the closed-loop generative design framework philosophy, where failed synthesis attempts in the SDL can feed back to retrain and improve the synthesizability predictor iteratively.

**Benchmarks for Generative Design of Inorganic Materials**

Evaluating generative models for inorganic materials remains an evolving challenge, as traditional metrics used in predictive modeling do not directly translate to generative tasks.

One recent approach is the S.S.U.N. (Symmetric, Stable, Unique, Novel) metric, introduced by WyFormer[76] to quantify the fraction of generated crystal structures meeting all four key criteria. It builds on prior advances: incorporating CDVAE[61] physics-driven emphasis on stability and invariance, extending MatterGen's "S.U.N."[77] focus (stable, unique, novel outputs) with an explicit symmetry requirement, and aligning conceptually with symmetry-centric frameworks like WyCryst[78], CrystalFormer[55] and SymmCD.[79] Each component addresses a fundamental quality: symmetric enforces crystallographic validity by requiring non-trivial space-group symmetry in the generated structure, (meta)stable[80] upholds physical realism by favoring energetically viable (potentially synthesizable) configurations, and unique and novel together promote generative diversity by ensuring outputs are distinct from one another and from known materials. By design, S.S.U.N. provides a useful framework for evaluating generative outputs in terms of crystallographic validity, stability, and novelty. However, such metrics remain imperfect proxies for real-world applicability, particularly as they do not fully capture synthesizability or functional performance under operating conditions.

To address these challenges, generative design frameworks may benefit from incorporating physics-informed foundation models capable of capturing multiple properties across diverse material classes. The integration of pre-trained MLIPs and multi-fidelity datasets can enable efficient screening and validation of generated candidates prior to experimental testing. Such approaches aim to improve both the quality and relevance of generated materials, though robust benchmarking standards across the field are still needed.

***Synthesis and Property Testing***

Central to any generative design framework is the availability of high-quality datasets, generated through both high-throughput computation and experiment. This is where AI integration with data-driven labs (or their digital twins) becomes crucial. Materials Acceleration Platforms (MAPs) and/or Self-driving laboratories (SDLs), equipped with robotics and automated characterization tools, rapidly test AI-generated predictions.[81] These platforms are increasingly viewed as essential components of closed-loop materials discovery, enabling rapid validation and iterative improvement of generative models, enabling the exploration of out-of-equilibrium materials by accessing synthesis conditions not easily captured in traditional workflows. The resulting data are fed back into the AI model, creating a self-correcting system that improves with each iteration. Challenges include the execution of dynamic test and validation tasks, encompassing planning through optimization algorithms, scheduling, and monitoring, as well as recovery and data collection. MAPs/SDLs have been implemented with great success in several applications. Li *et al.* conducted 250 autonomous experiments to discover optically active chiral perovskite nanocrystals, significantly faster than conventional methods.[82] Burger *et al.* performed 688 experiments over eight days of continuous operation to identify an optimal photocatalyst formulation for hydrogen production from water.[83] The identified catalyst was six times more active than previously known formulations, demonstrating the SDL's ability to rapidly explore and optimize chemical space.

Additionally, MAPs and SDLs have proven effective in optimizing material properties, such as thin-film synthesis. MacLeod *et al.* identified low-temperature synthesis conditions for palladium films, reducing the temperature by 50°C compared to prior methods.[84] Another SDL reduced the number of experiments required for optimizing 3D-printed structures by 60-fold. Inorganic materials present additional challenges for high-throughput experimentation. Unlike organic compounds, which are often synthesized using solution-based chemistry, many inorganics require high-temperature reactions or complex thin-film deposition techniques. While automation of these methods is still developing, advances in robotics for solid-state chemistry and AI-driven optimization algorithms show promise, guiding experimental iterations by selecting the most promising materials for synthesis based on prior results.[85] Despite these advances, significant challenges remain in scaling autonomous experimentation for inorganic materials, particularly for systems requiring high-temperature synthesis, complex processing conditions, or multi-step fabrication workflows.

Overall, the experimental data generated by autonomous/self-driving labs should be fed into foundational models for active learning. An emergent recent area is the possibility of enabling this through Agentic AI driven workflows. Typically, the bottleneck is in the generation of *in vivo* digital twins that capture the complexity of the particular problem statements for each multi-property domain. Surrogate models can be built upon well-sampled experimental data in the lab, which can then be run through computations and therefore the validation can be completed efficiently. The use of LLMs for in-context learning[86] to supplement existing active learning algorithms such as Bayesian optimization[87–89] and reinforcement learning[90] is an exciting frontier in this direction. Recent self-driving laboratories increasingly treat experimentation as a policy-driven, agentic workflow in which autonomous hardware, on-instrument analytics, and model-assisted planning are orchestrated end-to-end. In inorganic materials, A-Lab integrates ab initio phase-stability screening, literature-trained natural-language models that propose solid-state synthesis recipes, robotic execution, and ML analysis of X-ray diffraction, with active learning that closes the loop by proposing improved follow-up recipes from observed outcomes and failures.[3] In chemistry, mobile-robot platforms have demonstrated closed-loop synthesis–analysis–decision cycles that fuse orthogonal measurements (for example, NMR plus LCMS) into an autonomous human-like decision-maker, triggering downstream steps without human input.[91]

At the software layer, foundation-model agents are now being explicitly used as the "controller" that binds literature retrieval, tool use, and lab actions: ChemCrow couples an LLM to expert chemistry tools and can autonomously plan and execute experimental tasks[92], and LLM-RDF decomposes reaction development into a multi-agent pipeline that interfaces with automated platforms via natural language.[93] These systems are enabling the lab to a continuously generating "ground truth" stream for model improvement in the regions of chemical/material space it can most efficiently sample. The remaining scaling bottleneck is constructing *in-vivo* digital twins that faithfully and invertible map controllable conditions to coupled, multi-property observables (structure/phase, microstructure, kinetics, performance), motivating bidirectional theory-experiment loops with uncertainty-aware stopping conditions[93], although these are still narrowly applied to specific domains and infrastructure. Ultimately, the integration of generative models with autonomous experimentation represents a key step toward closing the loop between prediction and realization, though practical implementation at scale remains an open challenge.

Next, we highlight several application domains that serve as illustrative examples of how a generative design framework may be applied in practice.

## Use Cases

### (a) Materials for green hydrogen production:

Green hydrogen production presents a materials grand challenge due to the need for advanced materials that can facilitate efficient and sustainable hydrogen generation.[94,95] One promising method, electrolysis, where water is split into hydrogen and oxygen using electricity, demands materials capable of driving the process with minimal energy loss, must resist corrosion in an aqueous environment, and must remain effective over long periods to make green hydrogen a viable alternative to fossil fuels.[96]

In this area, a foundation model could be used to design and predict new materials, identifying efficient and cost-effective electrocatalysts that can outperform traditional platinum[97] or iridium-based systems.[98,99] Additionally, the inverse process of hydrogen burning in fuel cells is also extremely challenging as currently it is fully reliant on platinum catalysts. High cost and high density of platinum will have a crucial effect on the availability and affordability of hydrogen vehicles.[98] Thus, there is an impending need for highly efficient, cost-effective (Pt-free) catalysts.

A promising alternative are high-entropy, two-dimensional (2D) materials. The possibility of tuning of the specific crystallographic phase (which can define metallic of semiconducting nature of the grown crystals) as well as the controllable introduction of the individual atomic defects and impurities for single atom catalysis may enable broad tuning of the catalytic performance of such structures. Concomitantly, the possible phase space of such crystal structures is enormous, which essentially calls for the use of autonomous laboratories as well as FM models. The FM has been designed to account for single atom defects through a new representation (Figure 2B), and experimentally, alloying of 2D materials has been reported. However, to achieve good electrocatalytic performance, precise control of the composition, crystal phase and defects in high quality 2D materials are required. While solution processing and exfoliation have given good results in the past, they are not good approaches to explore a larger chemical space combinatorically. Physical deposition techniques such as molecular beam epitaxy (MBE), chemical vapour deposition (CVD) or metal-organic CVD are more suited for these and have been extensively used in the deposition of thin and thick films. Therefore, an AI driven CVD/MBE platform with optimization on the loop would represent a promising autonomous laboratory setup for this application (Figure 3). In a generative design framework, the predictive design of high-entropy 2D materials can be coupled with an active learning setup for automated growth, integrating growth and electrocatalytic performance into a single closed loop implementation.

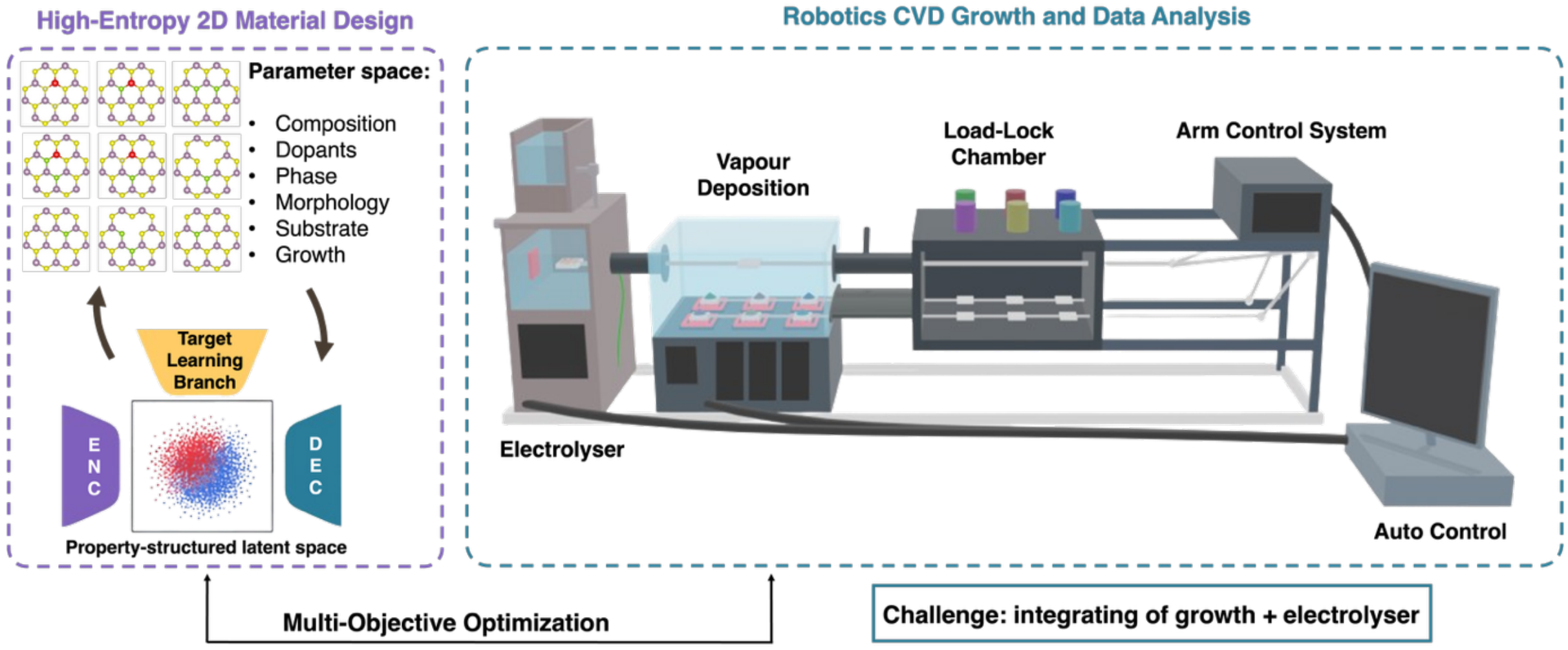

**Figure 3. Generative design framework for green hydrogen materials discovery**. The framework combines a generative model with robotic CVD synthesis and catalysis data collection. Structure of experimentally synthesized materials can be obtained by integrating an algorithm with the outputs of the robotic system. Multi-objective optimization is used to navigate a complex design space, accelerating discovery and feedback-driven improvement.

### (b) Materials for thermal barrier coatings:

Enhanced engine efficiency in aircrafts could be achieved by increasing the engine's operation temperature. The state-of-the-art Thermal Barrier Coating (TBC) material, yttria-stabilized zirconia (YSZ), is approaching its temperature limit, due to the destructive tetragonal-to-monoclinic phase transformation when temperatures exceed 1200°C. Advancing TBCs is a multi-property optimization problem where the target material needs to outperform YSZ by having a lower thermal conductivity, higher structural and phase stability, better adhesion strength while maintaining the desirable stiffness and compatible thermal expansion coefficient.[100] High entropy materials (HEM), including high entropy oxides, offer a materials-design route for tuning the relevant properties over a wide compositional range.[101] In particular, the increased configurational entropy, arising from arrangements of the 4 or more substitutional elements and defects, (i) promotes the formation of a single solid-solution phase and (ii) lowers the thermal conductivity through the increased scattering of phonons arising from disordering. DFT[102,103] provides an avenue for high-throughput (HT) pre-screening *via* computational means, where reliable formation enthalpies and thermal properties can be derived.[100] For the latter, the DFT-derived harmonic and anharmonic force constants enable accurate predictions of the lattice thermal conductivities via solving of the Boltzmann transport equations.[104–107] However, such high-fidelity methods become intractable when large supercells are required as in the case of HEMs. In these cases, MLIPs becomes essential to speed up the computations of the energetics and forces of the material structure.[108] In instances where forces are not required, such as in the evaluation of solid-state phase diagrams in crystalline substitutional alloys/compounds, cluster expansion (CE) is a suitable surrogate model for the rapid computation of the formation energies of millions of atomistic configurations on a fixed lattice.[109–111] This enables the construction of order-disorder phase stability diagram across composition *via* Monte Carlo simulations.[112,113] Moreover, the short-range order (SRO) within the solid-solution phase could be reliably predicted too.[41,75] Coupled with theories such as Debye-Callaway approach[114], (lattice) thermal conductivities could be predicted reliably and rapidly even for large unit cells. Automating the above workflow allows the construction of a large HEM dataset that relates structure to thermal properties.

Efficient film deposition must balance high material quality with speed to explore large chemical spaces. Combinatorial methods like pulsed layer deposition or molecular beam epitaxy enable this, especially when paired with high-throughput characterization tools such as *in-situ* temperature-dependent XRD for thermal stability and expansion, and automated XPS scans for rapid compositional analysis, greatly accelerating materials discovery. Once the model is trained with a combination of theoretical and experimental data, it can be used for the inverse design of synthesizable HEM compositions with optimal values of (high-temperature) stability, thermal conductivity, thermal expansion, adhesion strength and hardness. Importantly, the high-throughput experimental data will further augment the HEM dataset, therefore enhance the robustness of the model. Therefore, a high-throughput thin film deposition and characterization platform, driven by optimization algorithms would be a good autonomous lab for this application (Figure 4). Again, within the context of a generative design framework, the integration of multiple computational tools combined with pulsed laser deposition and property evaluation could provide a closed loop bridging experiment and theory.

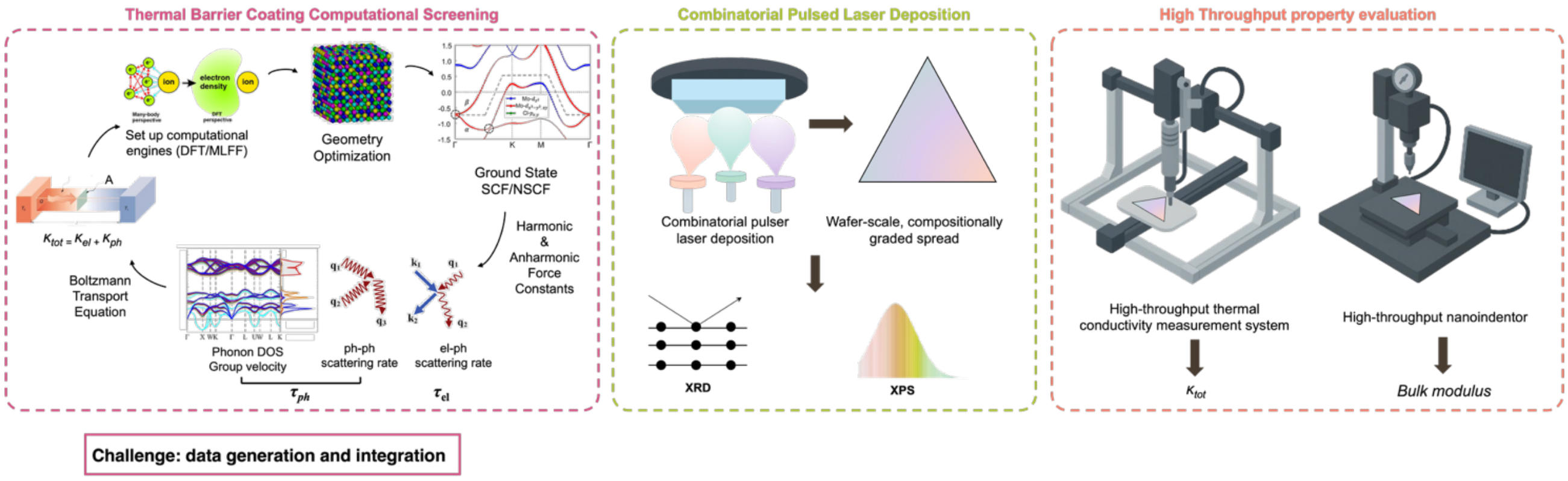

**Figure 4. Generative design framework for thermal barrier coating discovery.** MLIPs enables curation of large datasets of high-entropy materials consisting of representative structures and their thermal properties across a wide compositional space. The dataset is augmented by high-throughput experimentations which also serve as a validation source. The multi-fidelity AI model trained on the combined datasets could be used to generate candidate compositions with optimized properties and explore new compositional design space *via* active learning.

### (c) Materials for quantum technologies:

Single photon emitters (SPEs) are pivotal for quantum technologies, yet they pose a grand challenge because they demand materials capable of consistent, precise photon production at ambient conditions for technological viability. Two-dimensional materials have emerged as a prolific platform for hosting SPEs[115–118] due to their intentional creation in situ during the growth of monolayer films.[119] The growth conditions, such as reactor pressure or a carrier gas, affect the optical response of the SPE-hosting films.[120] The mechanisms of the defect formation can be elucidated at a level of DFT,[121] exemplified by the formation of carbon impurities in carbon-doped hexagonal boron nitride (hBN:C).[122] As illustrated in Figure 5A and 5B, carbon substitution for nitrogen (CN) and for boron (CB) constitutes simple acceptor-like and donor-like impurity states, respectively. When combined into the CNCB dimer defect, the midgap states transform into molecular-like occupied and empty orbitals. The relocation of the electron from the donor-like CN level to the occupied molecular-like CNCB level relaxes a substantial amount of energy (several eV), stabilizing complex defect structures in the hBN lattice. The defect formation mechanisms can drive the growth of hBN films hosting different defects at controllable concentrations. Technologically viable SPEs must exhibit appropriate photonic performance indicators (Figure 5C). These include high brightness and purity of quantum emission. Different applications require specific spectroscopic features, such as radiative resonances at exact wavelengths exhibiting narrow linewidths. The photophysical properties of defect-based SPEs are described within the Franck-Condon model, which considers optical transitions between electronic levels with simultaneous emission/annihilation of phonons and lattice reconstruction in the excited state. The accurate predictions of the optical response require advanced ab initio modelling, such as variations of GW approaches and quantum embedding.[123–126] These are computationally complex and expensive, preventing the creation of meaningful defect databases, which include information on the optical properties of SPEs.

Based on these considerations, an example of autonomous[127] SPEs synthesis platform is suggested, illustrated in Figure 5D. The growth of hBN:C films and optical characterization are coupled by a robotic arm moving the wafers from the growth chamber to the microspectroscopy system. The process is autonomously driven by artificial intelligence, which conducts confocal mapping of the optical response of the grown films, identifying the spectral features characteristic of SPEs and performing photon correlation experiments for verification. Based on this data, a reward function is constructed to quantify the density of optically active defects concurrently with their brightness and purity of quantum emission. Such a multimodal reward function needs to be optimized via machine learning by exploring the multidimensional space of growth parameters. The decision-making in traversing this parameter space should be driven by the knowledge about the defect formation mechanisms embedded into the foundational model via the defect structure database. The best samples produced via such a procedure are selected for advanced testing, which involves the identification of atomic and electronic defect structures via transmission electron microscopy[128] and scanning tunnelling microscopy[129] combined with advanced optical characterization. These results, interpretable within *ab initio* models, may provide a complete understanding of the photophysical properties of SPEs with known defect structures. A foundational model, iteratively enriched with the structure-property relations, will synthesize SPEs with desired properties. Such SPE-hosting films can be afterwards integrated into photonic structures and devices[130–133], offering avenues to satisfy the requirements of quantum telecommunication, computing, and sensing**.** In this example, integrated approaches could connect the atomic scale to electronic structure and optical properties, determined by defect engineering by design, linking again to experimental measurements.

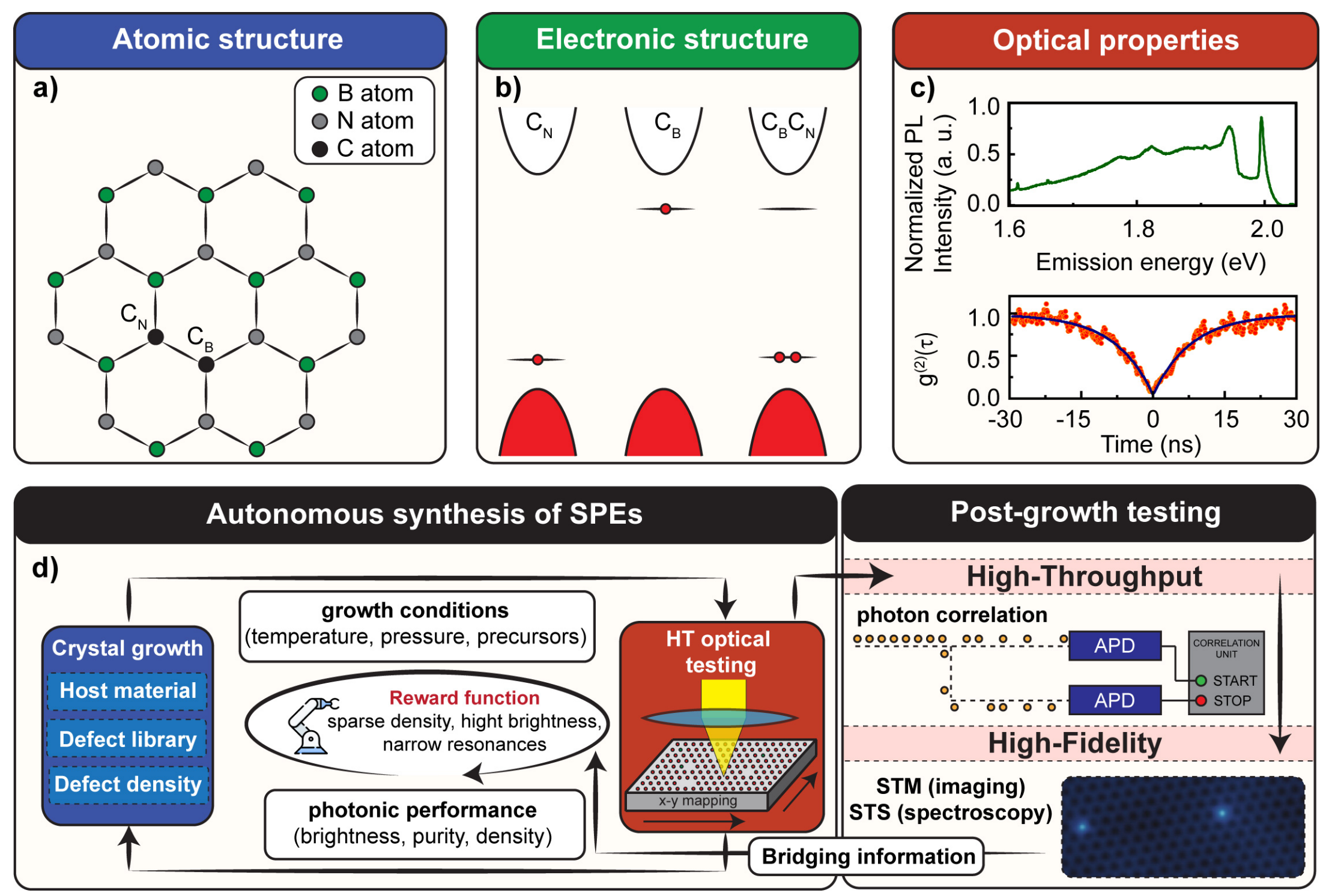

**Figure 5. Generative design framework for single photon emitters (SPEs).** a) Example defect in carbon-doped hBN. b) Schematic of acceptor- and donor-like states from carbon substitutions and molecular-like states from dimers. c) PL spectrum and second-order photon correlation confirming quantum emission. d) AI-driven autonomous CVD workflow with optical testing and feedback via a reward function (based on brightness, defect density, purity). ML and a foundational model guide synthesis, with top samples undergoing advanced testing to refine the model and optimize growth parameters.

### (d) Materials for $CO_2$ reduction:

Electrocatalysis is a key technological pathway for converting $CO_2$ into value-added products. Traditionally, electrocatalysts have been synthesized using solvent-based methods or magnetron sputtering, which are largely limited to unary or binary compositions and hindered by low-throughput fabrication and maintenance challenges. As the complexity of catalyst compositions increases (e.g., ternary or higher-order systems), these limitations become more pronounced. Recent advances in HT synthesis methods now enable one-pot, multi-elemental catalyst fabrication at low substrate temperatures, facilitating direct integration into membrane electrode assemblies (MEAs) or flow cells.[134] A catalyst discovery framework, integrating generative design, HT validation, and active learning, could help address these challenges (Figure 6A). Notably, here the catalyst material transforms in the testing conditions and hence one can imagine the catalytic testing to be a 'controlled black-box' where the testing data is collected and fed into the multi-modal inverse design model. The data collected constitutes the catalytic material before and after the testing, thus enabling a closed-loop material design approach. To illustrate in more detail, initially, a generative model is trained on existing experimental and theoretical data to propose catalyst candidates within a defined chemical space, optimized for activity, selectivity, and stability (Figure 6B). HT fabrication, including combinatorial mixing and rapid sintering, is then used to synthesize these candidates, yielding catalyst GDEs directly suitable for cell test. HT characterization follows rapidly, providing structural and compositional data. Coupled with automated MEA testing and near-line gas/liquid chromatography analysis, this allows for a fast correlation between synthesis parameters, catalytic performance, and material properties. Validation results are fed back into an active learning loop to iteratively refine predictions and guide subsequent synthesis rounds. Moreover, mechanistic insights during operation can be obtained via in-line mass spectrometry or optical spectroscopy, enhancing model interpretability. Ad-hoc DFT calculations based on observed active sites or catalytic pathways further refine the generative model. This integrated generative design approach could enable iterative and more interpretable discovery of optimal $CO_2$ reduction electrocatalysts.

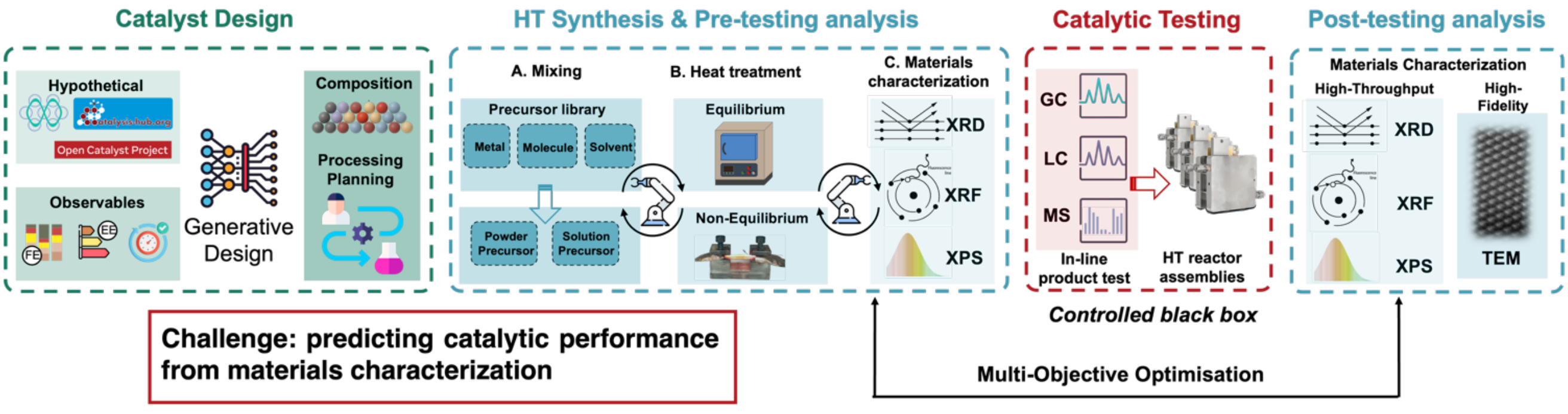

**Figure 6. Generative design framework for catalyst discovery.** (1) Generative design of catalysts; (2) HT-fabrication *via* diverse syntheses; (3) HT-catalyst optimization combining materials characterization, catalytic performance and mechanism evaluations, and Bayesian optimization.

## Outlook

A Generative design framework for inorganic materials offers significant potential to accelerate discovery and enable new classes of functional materials. Applications of such frameworks in sustainability, energy, and next-generation electronics could provide solutions to some of the most pressing global challenges, such as carbon neutrality and energy efficiency. These integrated approaches propose an iterative, closed-loop, and capable of continuous learning framework combining both simulations and experiments. However, realizing the full potential of such a framework requires coordinated progress on several interrelated fronts. Key examples include the development of truly generalizable FMs for inorganic materials that encode crystallographic symmetry and local bonding, and the integration of structural imperfections (e.g., vacancies) directly into generative processes to treat defects as fundamental design parameters. Furthermore, closing the synthesis-design loop requires advances in autonomous laboratories (e.g. reliable sample transfer), supported by open and interoperable platforms that standardize benchmarks like the S.S.U.N. metric, thus redefining discovery itself.

### Toward universal foundation models for inorganic materials

The vision for materials FMs is ambitious: models that are pre-trained on vast datasets of composition-structure-property relationships, adaptable to diverse downstream tasks via fine-tuning, and interpretable with respect to known chemical and physical principles. To achieve this, several milestones must be reached. First, representation schemes must evolve beyond stoichiometry or atomic coordinates to encode crystallographic symmetry, space group operations, site disorder, and local bonding environments in a manner amenable to machine learning. Symmetry-aware and equivariant models are promising in this regard, though they remain constrained by limited training data and the difficulty of representing long-range or defect-related phenomena.

Second, FMs must be trained to learn across data modalities and fidelities. Combining low-fidelity approximations with high-fidelity calculations and experimental measurements requires not just model design, but new approaches to uncertainty quantification and transfer learning among others. Moreover, FM training must extend beyond simple property prediction to include constraints of stability, synthesizability, and domain relevance. We anticipate the emergence of multi-task, multi-property models that can jointly learn and predict diverse physical properties, adapt to new material classes with minimal additional data, and support conditional generative design.

**Embracing imperfections as degrees of freedom for design**

A defining feature of real materials is imperfection: vacancies, substitutions, dislocations, grain boundaries, and surfaces among others. However, most generative models remain restricted to perfect crystals. A crucial shift is needed: from viewing defects as complications to treating them as fundamental design parameters. This requires not only new multi-level representation schemes, capable of encoding point and extended defects in machine-readable formats, but also datasets that capture the impact of these features on material properties.

Emerging methods, such as defect-aware graph networks and generative models conditioned on local disorder, hold promise for enabling the targeted design of functional imperfections. Applications include single-atom catalysts, defect-mediated quantum emitters, and doped semiconductors. The broader implication is that the materials design space becomes significantly richer when defects are treated not as noise to be corrected, but as levers for tuning performance. Integrating this perspective into generative design frameworks will be essential to unlocking new classes of high-performance materials.

**Plugging the gap: autonomous experimental validation**

Prediction does not equate discovery. For generative models to meaningfully accelerate materials innovation, their outputs must be experimentally validated. Autonomous laboratories provide a scalable solution, offering closed-loop systems where generative predictions guide synthesis and testing, and validation data refine subsequent model iterations. Realizing this potential will require further advances in robotic synthesis for inorganic materials, from solid-state synthesis to thin-film deposition and multi-elemental alloying, as well as high-throughput characterization techniques capable of capturing both structure and function. Critically, *for autonomous laboratories to narrow the gap between materials design and the discovery of functional systems, they must be built minimizing both sample transfer and geometric mismatch between the as-synthesized material and the form required for measurement*. Thermoelectric materials provide a salient example. State-of-the-art thermoelectrics are dense, bulk solids; accordingly, it is advantageous not only to synthesize materials directly in bulk form, but also to produce them in geometries that match those used for subsequent testing, for example, parallelepipeds for temperature-dependent electronic transport measurements or disks for thermal transport characterization. This approach yields a twofold acceleration: (i) synthesizing dense bulk materials obviates the need for post-synthesis densification, and (ii) producing samples in measurement-ready geometries eliminates the need for cutting or reshaping prior to performance evaluation.[135]

A good opportunity may lie in embedding domain-specific knowledge, such as phase diagrams, reaction mechanisms, or kinetic constraints, directly into synthesis planning. Agent-based systems that integrate this knowledge with real-time feedback and reinforcement learning algorithms could autonomously navigate vast experimental parameter spaces, identify optimal synthesis conditions and accelerating convergence toward target properties. Critically, the data generated from such systems must be standardized, labelled, and fed back into FMs to enable continual improvement. In this way, autonomous and self-driving laboratories become not just validation platforms, but of model refinement engines.

**Redefining the act of discovery**

Perhaps the most profound shift implied by the generative framework described in this work is conceptual. Traditionally, materials discovery has meant incremental optimization: finding a composition that outperforms a benchmark. The fusion of generative AI with physical constraints and data-driven experimentation (SDLs/MAPs) enables a distinct notion: the identification of materials that exhibit novel structure types, unexpected mechanisms, and emergent properties not anticipated by existing theories.

This perspective aims to reframe discovery not as a search over known possibilities, but as an expansion of what is thinkable within the constraints of chemistry and physics. As models learn to generalize across domains and propose candidates beyond human intuition, a window of opportunity opens to uncover new classes of materials defined not by prior knowledge but by learned representations.

Materials in Context

Technology maturity: Basic principles observed – the generative design framework is at an early research stage, with proof-of-concept demonstrations in a few domains.

Key challenges:

• Capturing and representing defects, disorder and multi-scale structures
• Integrating generative design with autonomous synthesis and characterization workflows
• Scaling training data and models for generalizable, transferable predictions
• Ensuring experimental reproducibility and reliability in closed-loop systems
• Addressing safety, sustainability and ethical considerations in AI-driven materials design
• Developing standardized benchmarks and evaluation metrics across materials domains

Potential impact: Once fully developed, these generative design frameworks could enable accelerated discovery of high-performance catalysts, batteries, thermoelectrics and other functional materials, reducing discovery times from years to days and supporting sustainable, circular materials economies.